\documentclass[10pt,conference]{IEEEtran}

\usepackage{amsmath, amssymb, bm, cite, epsfig, psfrag}
\usepackage{graphicx}
\usepackage{subcaption}
\usepackage[top    = 0.7in,
bottom = 0.95in,
left   = 0.7in,
right  = 0.7in]{geometry}
\usepackage{array}
\usepackage{textcomp} 
\usepackage{longtable}
\usepackage{supertabular,booktabs}
\usepackage{authblk}
\usepackage{multirow}
\usepackage[usenames,dvipsnames]{xcolor}
\usepackage{etoolbox}
\usepackage{pbox}
\usepackage{romannum}
\usepackage{enumerate}
\usepackage{colortbl}
\usepackage{url}
\usepackage[switch]{lineno} 
\usepackage{bm}
\usepackage{float}
\usepackage{fancyhdr}
\fancypagestyle{firststyle}{\fancyhf{}
\fancyhead[L]{Aditya Chopra, Andrew Thornburg, Ojas Kanhere, Abbas Termos, Saeed S. Ghassemzadeh, and Theodore S. Rappaport ``Real-time Millimeter Wave Omnidirectional Channel Sounder Using Phased Array Antennas",\textit{ in GLOBECOM 2020 - 2020 IEEE Global Communications Conference, }Taipei, Taiwan, Dec. 2020, pp. 1--7}
}

\graphicspath{{figures/}}

\begin{document}
\title{Real-time Millimeter Wave Omnidirectional Channel Sounder Using Phased Array Antennas}	

\author[1]{Aditya Chopra}
\author[1]{Andrew Thornburg}
\author[2]{Ojas Kanhere}
\author[3]{Abbas Termos}
\author[1]{\\Saeed S. Ghassemzadeh}
\author[2]{Theodore S. Rappaport}
\affil[1]{AT\&T Labs, \{aditya\_chopra, andrew\_thornburg, 
		saeed\}@labs.att.com}
\affil[2]{NYU Tandon School of Engineering, \{ojask, tsr\}@nyu.edu}
\affil[3]{University of Notre Dame,\{atermos\}@nd.edu}

\maketitle
\thispagestyle{firststyle}
\begin{abstract}
Characterization of the millimeter wave wireless channel is needed to facilitate fully connected vehicular communication in the future. To study the multipath-rich, rapidly varying nature of the vehicular propagation environment, fast millimeter wave channel sounders are required. We present a channel sounder design capable of covering 360 degrees in azimuth and 60 degrees in elevation with 200 individual beam directions in 6.25 ms by using four phased arrays simultaneously. The channel measurements are accompanied by high resolution positioning and video data, allowing channel sounding to be conducted while either the transmitter, or the receiver, or both are moving. Channel sounding campaigns were conducted at multiple urban locations with light traffic conditions in Austin, Texas. Preliminary results show that beam selection at the receiver can lower the effective pathloss exponent to 1.6 for line-of-sight and 2.25 for non line-of-sight. 



\end{abstract}
    

\section{Introduction}\label{sec:Introduction}

Over the past few years we have witnessed an enormous amount of progress on standardization of fifth generation (5G) wireless communications leading to initial deployments of 5G networks using sub-6\,GHz and millimeter wave (mmWave) spectrum in many parts of the globe\cite{ATT_mmWave}. These networks are designed to enable billions of new connections with high data rates, ultra low latency, and high capacity. One particular use case of interest is vehicle-to-vehicle (V2V) or vehicle-to-infrastructure (V2I) communications in mmWave spectrum. With the wide bandwidth and ultra low latency, as offered by mmWave in 5G standards \cite{Rappaport_2013b}, ultra-reliable vehicular communication with rapid data exchange enables novel vehicular applications such as autonomous driving and platooning, collision warning, and vehicular localization\cite{Va_2016}. The next generation of wireless networks are expected to go further towards wireless cognition, by forwarding the computation of the huge quantities of sensor data to edge servers \cite{Rappaport_2019}.

In order to achieve these ambitious goals, it is crucial to fully characterize the behavior of the wireless channel in a vehicular communication setting. One of the unique features of mmWave communications is the need for highly directional transmission and reception in order to overcome high propagation losses\cite{rappaport_2015}. For vehicular communications in particular, the signal strength in different directions can change rapidly due to the motion of the transmitter (TX), the receiver (RX), the relative height difference between the transmitter and receiver, and other objects in the local environment that act as blockers, reflectors, or scatterers. It is therefore important to conduct channel measurements that cover a variety of arrival, or departure directions and are fast enough to capture the rapid changes in the channel. 

A large amount of previous research on mmWave channel sounders and V2V channel measurements uses passive directional antennas. For example, the sliding correlator based channel sounder described in \cite{MacCartney_2015} uses high gain directional horn antennas, mounted on gimbals at the transmitter and receiver, that  are rotated in azimuth and elevation planes in order to measure the omnidirectional channel. However, due to the low speed of the gimbal, this sounder is not suitable for vehicular channel measurements. While the channel sounder described in \cite{Park_2018} is capable of measuring large and small scale channel parameters at $28\,\rm{GHz}$, it is also not suitable for vehicular measurements with small coherence time. The time domain sounder in \cite{Prokes_2018} operates at $59.6\,\rm{GHz}$ with bandwidth of $8\,\rm{GHz}$ utilizing omnidirectional antennas for sounding, which due to the lack of directional antennas, angular information of the mmWave channel could not be measured. Sliding correlation based channel sounders at $38\,\rm{GHz}$ and $60\,\rm{GHz}$ were also developed with an RF bandwidth of $1.9\,\rm{GHz}$\cite{Ben_Dor_2011}. Vehicular angle-of-arrival measurements were conducted in a parking lot using the channel sounder in\cite{Ben_Dor_2011}.  The mmWave phased array based channel sounder in \cite{Bas_2019} uses steerable beams to measure the wireless channel within $90^{\circ}$  sector at the receiver with a beam switching time of less than $2\,\rm{\mu s}$.
An omnidirectional sounder in $60\,\rm{GHz}$ band using fast arrays for beam sweeping was developed in \cite{Caudill_2019}, however, it only sweeps the beam in the azimuth plane and only indoor measurements were conducted. 

To meet the requirement for a vehicular $360^{\circ}$ channel sounder with fast beam switching, a novel Real-time Omni-Directional Channel Sounder (ROACH) was developed by AT\&T Labs to conduct channel sounding measurements in V2V and V2X environments.

The remainder of this paper is organized as follows. Section \ref{sec:CS_design} describes the channel sounder design. System calibration and verification procedures are given in Section \ref{sec:calibration}.  Channel measurements and preliminary results are provided in Section \ref{sec:measurements} followed by conclusions and future work in Section \ref{sec:conclusion}.

\section{Channel Sounder design}\label{sec:CS_design}
ROACH was developed as a real-time wideband correlation type channel sounder \cite{silva_2018}. Such channel sounders operate by transmitting a known signal $s(t)$ that typically has an autocorrelation function $s(t) * s(t)$ equal to the Dirac-Delta impulse function $\delta(t)$. Given a channel impulse response $h(t,\tau)$, and transmit signal $s(t)$, the received signal due the channel only, is a convolution $r(t) = h(t,\tau) * s(t)$. Upon correlating $r(t)$ with $s(t)$, the receiver system can extract the estimate $h(t,\tau)$ of the wireless channel. Real-world impairments such as additive Gaussian noise, IQ impairments, frequency and sampling rate offset, and amplifier non-linearities can corrupt the channel estimates. A good channel sounder design must either correct or limit the impact of such impairments through proper selection and operation of system components, sounding waveforms, and digital algorithms. 

\subsection{Transmitter}\label{subsec:tx}
A common transmit sequence to use in wideband channel sounders is the Zadoff-Chu (ZC) sequence\cite{hua_2014}. The complex value $x_u[n]$ of a length $N_{ZC}$ root ZC sequence with parameter $u$ is given by

\begin{align}
  &x_u[n]=\text{exp}\left(-j\frac{\pi un(n+1)}{N_\text{ZC}}\right), \\
  \text{where}~ &0 \le n < N_\text{ZC},  \nonumber \\ 
  &0 < u < N_\text{ZC}~\text{and}~\text{gcd}(N_\text{ZC},u)=1  \nonumber
\end{align}

The ZC sequence has an autocorrelation of $N_{ZC}\delta[n]$, and constant amplitude in time and frequency domain\cite{hua_2014}. Its constant amplitude property is especially useful in helping transmit the sequence at high power levels with low risk of inducing power amplifier non-linearities. The sounding signal is generated by passing the sequence $x_u[n]$ through a RF front-end at a sampling period of $T_s$, consequently 

\begin{equation}
	s(t) = x_u[(k\,\text{mod}\,N_\text{ZC})T_s]\,\forall\,k\,\in \mathbb{N}, t=kT_s
\end{equation}

\begin{figure}[t]
	\centering
	\includegraphics[width=0.46\textwidth]{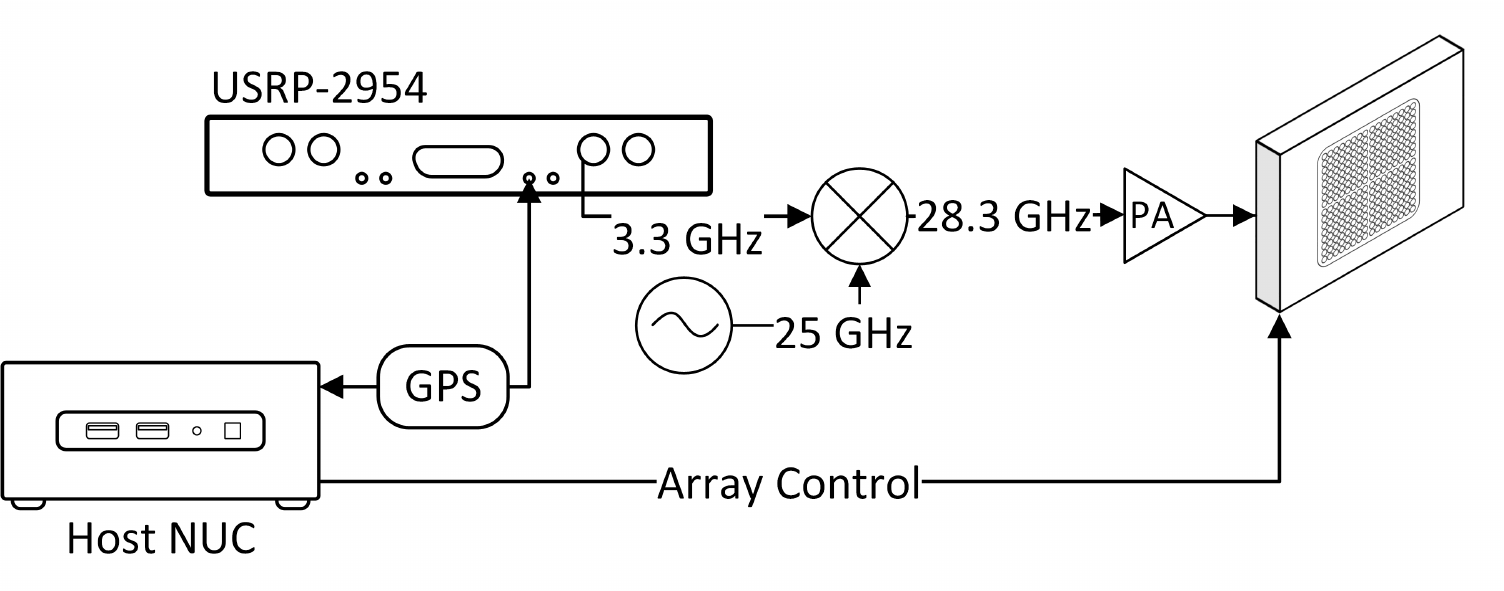}
	\caption{Block diagram of channel sounding transmitter system. }
	\label{fig:tx_block_diagram}
\end{figure}

Fig. \ref{fig:tx_block_diagram} illustrates the ROACH transmitter block diagram. It uses National Instruments' USRP-2954 software defined radio (SDR) to generate and transmit a ZC sequence at any chosen sampling rate $\frac{1}{T_s}<160\,\rm{MHz}$. Since the USRP-2954 transmitter supports an intermediate frequency (IF) between $1-6\,\rm{GHz}$, the IF signal can be upconverted by either $25\,\rm{GHz}$ or $37.5\,\rm{GHz}$ for the RF output signal to occupy the $28\,\rm{GHz}$ or $39\,\rm{GHz}$ mmWave band, respectively.

The output of the upconverter is fed into a power amplifier (PA) having $44\,\rm{dB}$ gain and $P_{1dB}$ compression point of $17\,\rm{dBm}$. Finally, the output of the amplifier is fed into a Anokiwave AWA-0134 Active Antenna Innovator’s Kit. The AWA-0134 is a 256-element phased array module operating in the $27.5$ to $30\,\rm{GHz}$ band. The beam direction in this phased array is controlled via Ethernet from a host computer inside the ROACH transmitter module. The beam width can also be controlled via selecting one of four beam types with narrow or wide widths. The allowed beam widths and their corresponding boresight gains are shown in Table \ref{table:awmf0134beams}. The boresight gain is the overall module gain, which is a combination of the gains from beamforming and Power Amplifiers on the module.

\begin{table}[h!]
\centering
\caption{Measured performance characteristics of the various available beams in the 256 element AWA-0134 Phased Array Module.}
\label{table:awmf0134beams}
\begin{tabular}{ | c | c | c| } 
 \hline
 Beam Type & 3dB Beam Width ($^{\circ}$) & Boresight Gain (dB) \\ [0.5ex] 
 \hline
 1 & 7.0 & 59.1 \\ 
 2 & 25.0 & 41.3  \\
 3 & 54.1 & 36.8  \\
 4 & 80.0 & 33.4 \\ 
 \hline
\end{tabular}
\end{table}

\subsection{Receiver}\label{subsec:rx}
\begin{figure}[t]
	\centering
	\includegraphics[width=0.46\textwidth]{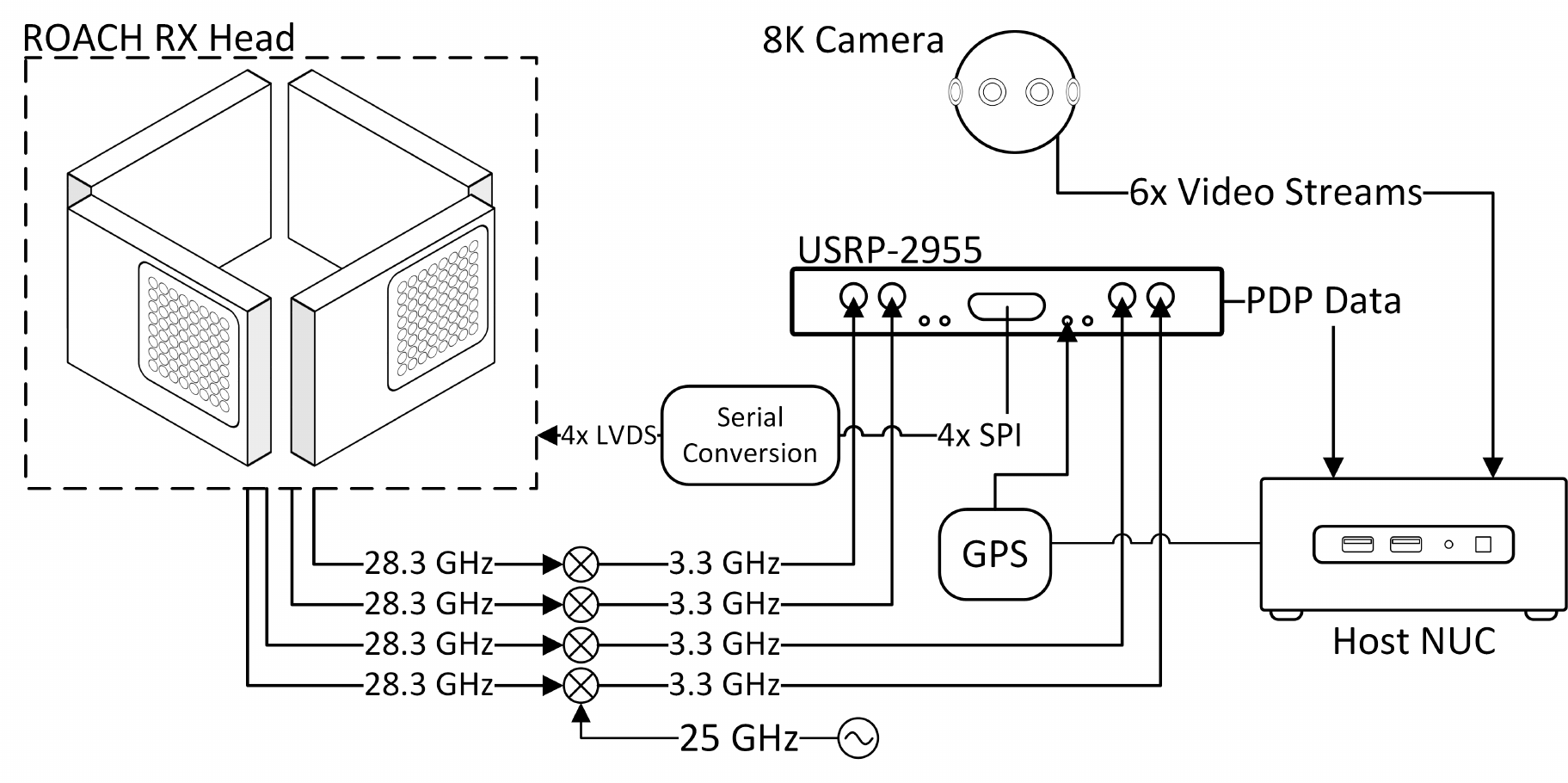}
	\caption{Block diagram of channel sounding receiver system. }
	\label{fig:rx_block_diagram}
\end{figure}

We now describe the ROACH receiver shown in Fig. \ref{fig:rx_block_diagram}. It is primarily comprised of four Anokiwave AWMF-0129 Active Antenna Innovator's Kit phased arrays, a multichannel RF to IF downconversion system, a National Instruments USRP-2955 four channel SDR, and an Intel NUC for startup and control, data collection, and user interface. 

The sounding signal is received through the AWMF-0129, which is a 64-element planar phased array module operating between $27.5$ and $30\,\rm{GHz}$. The receive beam can be swept between $\pm60^{\circ}$ from boresight in azimuth and $\pm 45^{\circ}$ in elevation. For omnidirectional reception in the azimuth, four AWMF-0129 arrays are placed in a square pattern shown in Fig. \ref{fig:rx_block_diagram}, each covering a one quarter sector of the azimuthal plane. Consequently, the beam sweeping range for each array is chosen to be $\pm 45^{\circ}$ in azimuth, and $\pm 30^{\circ}$ in elevation from boresight. 

The ROACH receiver can accept a user-defined beam sweeping codebook of up to $1024$ beams per array sector. The receiver cycles through the beams at user defined intervals that are synchronized such that the capture interval length is exactly an integer multiple of the sounding signal duration used for averaging. Higher averaging yields better estimates of the channel measurements but come at a cost of longer omnidirectional scan durations, a tradeoff typically seen in most channel sounder designs. Each beam in the user-defined codebook contains beam direction information as well as the beam width information. The AWMF-0129 array also allows for four beam width options, these options along with their corresponding boresight gains are listed in Table \ref{table:awmf0129beams}. The boresight gain is the overall module gain, which is a combination of the gains from beamforming and Low-Noise Amplifiers on the module. The same codebook is used to sweep the coverage sector of each of the four arrays.

\begin{table}[h!]
\centering
\caption{Measured performance characteristics of the various available beams in AWMF-0129 Phased Array Module.}
\label{table:awmf0129beams}
\begin{tabular}{ | c | c | c| } 
 \hline
 Beam Type & 3dB Beam Width ($^{\circ}$) & Boresight Gain (dB)\\ [0.5ex] 
 \hline
 1 & 14.2 & 47  \\ 
 2 & 16.8 & 43.3  \\
 3 & 18.7 & 34.3  \\
 4 & 16.5 & 30.3 \\
 \hline
\end{tabular}
\end{table} 

The AWMF-0129 can switch beams via Ethernet or serial-parallel interface (SPI) control. While Ethernet control requires more than $1\,\rm{ms}$ for beam switching, the SPI control interface can accomplish beam switching in less than $1\,\rm{\mu s}$, depending on the user designed SPI clock frequency. We used the NI USRP-2955 general purpose input output (GPIO) subsystem to control the four phased array receivers. A custom digital conversion box was made to convert single-ended 3.3V transistor-transistor logic (TTL) output of the USRP GPIO to four separate synchronized low voltage differential signal (LVDS) digital inputs for each of the phased arrays. Our SPI implementation in the USRP can preload beam coefficients on the array within $60\,\rm{\mu s}$. A dedicated latch signal is used to command the array to switch after the beam is loaded. The switching time of the array is less than $1\,\rm{\mu s}$. This acts as the lower limit on the duration our sounding signal, in order to switch the arrays to the next beam in the codebook after capturing a single instance of the sounding sequence.

The received signals from each of the four arrays are fed to a four-input down-converter with a $25\,\rm{GHz}$ local oscillator producing four simultaneous IF signals. The received IF signals are captured by the USRP and subsequently cross-correlated with the ZC sequence to generate the power-delay profiles (PDPs) and received correlated power. The signal processing algorithms are executed in real-time on a field programmable gate array (FPGA) inside the USRP. The channel measurements are timestamped and sent to the host computer via high speed interface for visualization and storage on high speed solid-state drives.

\subsection{Additional Sensors}
The ROACH system is also equipped with GPS receivers for high-accuracy data logging at both the transmitter and receiver. The model of the GPS receiver is the REACH M+, manufactured by Emlid Ltd. \cite{emlid}. This receiver can operate at 14Hz, or a GPS measurement every 70 ms. It can also apply Real-Time Kinematics (RTK) correction in order to achieve sub-$1\rm{m}$ space positioning accuracy \cite{emlid}. The GPS receiver was connected to the internet via a high-speed LTE hotspot, and the RTK corrections were provided via a service subscription from SmartNet NA Inc.\cite{smartnet}. This allows ROACH to conduct V2I and V2V channel sounding measurements where either the transmitter, or the receiver, or both are mobile. The GPS signals were also used to discipline the oscillators on both the transmitter and receiver within 10 parts per billion\cite{usrp}.

To further augment channel data collection, a six lens $360^{\circ}$ 8K resolution camera is mounted at the center of the ROACH receiver, capable of capturing video at 60 frames per second. This provides useful visual feedback of the clutter in the environment while experiments are conducted. The video is recorded and can be used in post processing to augment the data visualization.

The ROACH system software contains novel algorithms to synchronize channel sounding, GPS data collection, and video logging. This allows the user to correlate channel measurements with the position of the transmitter and receiver, as well as the environment around the receiver as seen by the $360^\circ$ video, with a high degree of accuracy. For example, the system can track the reflection off a moving car near the receiver.

\subsection{System Parameters}\label{subsec:params}
In all experiments described within this article, the value of $N_{ZC}$ is set as $8192$, and $u$ is set as $1729$. The output sampling rate of the signal in the USRP-2954 is set to $65.536\,\rm{MHz}$, which combined with the chosen value of $N_{ZC}$ results in the time duration of a single sounding sequence being equal to $125\,\rm{\mu s}$. Typical $3^{\rm{rd}}$ generation partnership program (3GPP) New Radio (NR) deployments in mmWave using a sub-carrier spacing of $120\,\rm{kHz}$ have a slot duration of exactly $125\,\rm{\mu s}$\cite{3GPP.38.211}. The settings used in ROACH were chosen to yield a sounding sequence length equal to a typical 3GPP NR slot duration in order to better relate channel measurements with NR system design parameters.

The sounding interval of $125\,\rm{\mu s}$ is sufficiently long for our phased array digital control modules to switch the beam every sounding instance. An azimuthal spherical segment was generated with the top and bottom planes defined by $\pm45^{\circ}$ elevation planes, respectively. Considering the trade-off between fine angular resolution and short sweep time required, this segment was tessellated by a $200$ cell uniform hexagonal lattice with the beam directions at the centers of each hexagon. The resulting spherical segment and tessellation pattern can be seen in Fig. \ref{fig:bp_quantized}. The beam directions were chosen such that the edges of the different array spans mesh together without any gaps. Each of the array sectors was assigned 50 of the 200 overall beams as its codebook, ensuring that there is no gap between the edge of the hexagonal cells between two different array sectors. With averaging set to $1$, the beams are switched every sounding interval of $125\,\rm{\mu s}$, resulting in an overall segment scan time of 6.25 ms. Each beam in the receiver codebook was assigned beam Type 2 from Table \ref{table:awmf0129beams}. The Type 2 beam was chosen because of its strong suppression of sidelobes at a small expense of increased 3dB beam width compared to the Type 1 beam.
 
The IF frequency was chosen as $3.3\,\rm{GHz}$, and the signal was upconverted by $25\,\rm{GHz}$, for a RF output frequency of $28.3\,\rm{GHz}$. The transmitter system was set up to point in boresight direction, i.e. azimuth and elevation set to $0^{\circ}$. The transmitter beam type was set to Type 3, allowing for a large coverage area at the cost of array gain.

\section{Sounder Calibration and Verification}\label{sec:calibration}
\subsection{Calibration}
In this sub-section, we describe the calibration procedure for the ROACH transmit and receive systems. The ROACH transmitter was calibrated at the point of the AWA-0134 phased array input using a Rohde and Schwarz NRP-Z11 power meter as the calibration reference. The maximum safe RF input level into the phased array was set as $-12\,\rm{dBm}$, which after accounting for cable losses ensured that the pre-amplifier was always operating in its linear region.

For the ROACH receiver system, the calibrated transmit module was connected via a Wilkinson divider to the output plane of the AMWF-0129 phased array. The other port of the Wilkinson divider was connected to the power meter. Using this setup, the transmit module would emit a sounding signal and the receive system would detect a single strong peak. Since the sounding signal has constant instantaneous power, the measured peak value was calibrated against the power meter reading. Note that this calibration was performed at high power levels that were within the range of the power meter specifications. In order to ensure accuracy at low power levels, no analog front-end settings were changed after calibration. Thus, lower power level measurements were purely derived digitally through the resolution of the ADC, requiring no further calibration.

The USRP radios are manufacturer calibrated to correct for IQ imbalances, and the frequency offset between the transmitter and receiver was maintained at acceptable levels by disciplining local oscillators using GPS timing signals.

The AWMF-0129 and AWA-0134 antenna arrays were calibrated in an anechoic chamber by the manufacturer and the beam pattern data was used for post processing of collected channel measurements.

\subsection{Verification}
To validate our calibration procedure of the ROACH system, over-the-air (OTA) verification is conducted in an anechoic chamber. The TX and RX phased arrays were placed $17\,\rm{ft.}$ apart with the TX and the front-facing RX phased arrays in boresight, aligned via laser. Due to the short separation, the transmit power was lowered in order to prevent damage at the receiver phased array. The channel measurements computed at the receiver were verified to be in agreement with calculated pathloss of $75.67\,\rm{dB}$ at $28.3\,\rm{GHz}$. This was repeated across all phased-arrays in the receiver.

\begin{figure}
	\centering
	\includegraphics[width=0.45\textwidth]{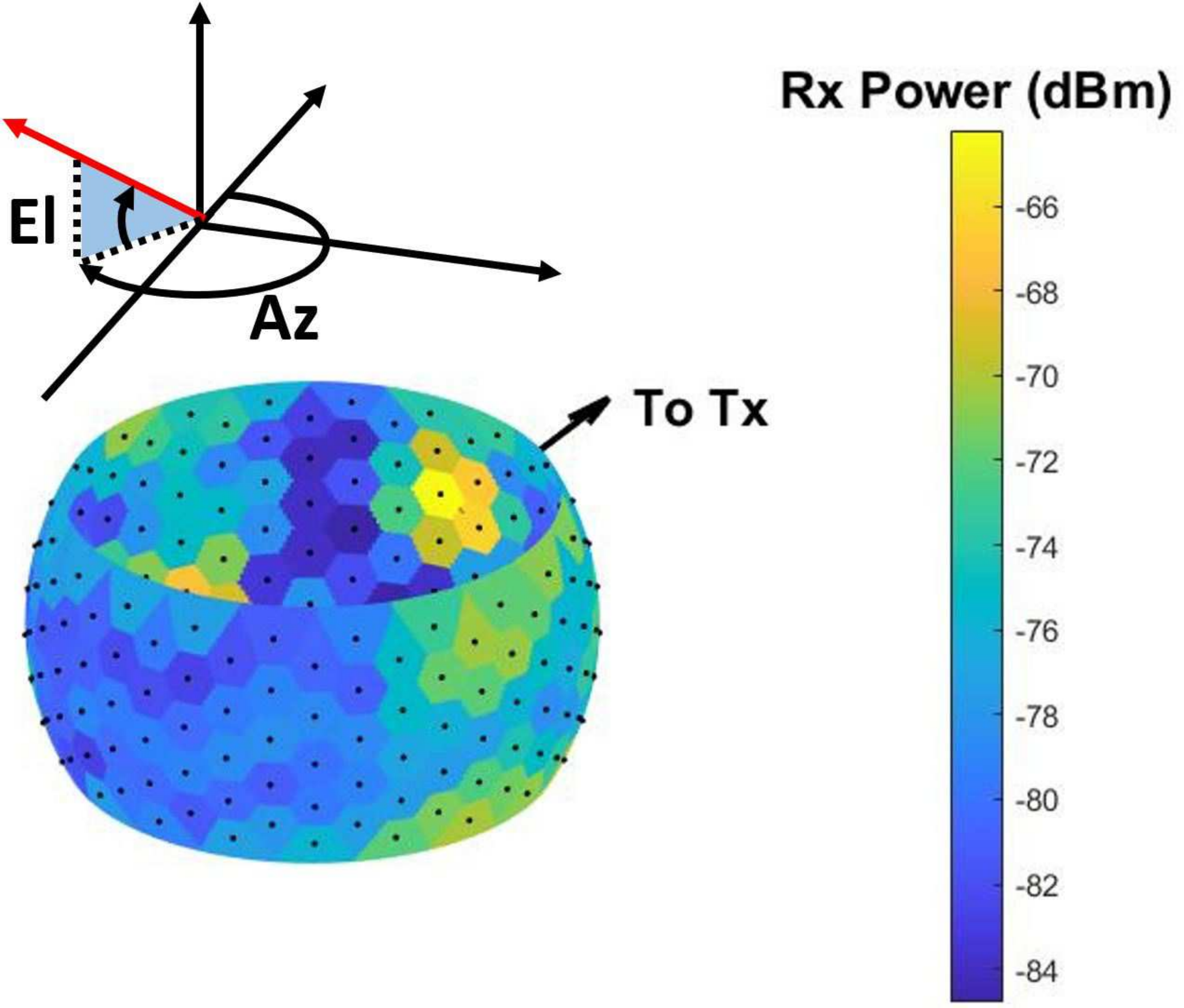}
	\caption{A heatmap of the received power across the 200 beams covering $360^{\circ}$ in azimuth and $60^{\circ}$ in elevation using the four RX phased arrays. The transmitter and receiver are placed in an anechoic chamber with the transmitter location shown by the arrow.}
	\label{fig:bp_quantized}	
\end{figure}

In addition, the RX beams of the four phased arrays were simultaneously electronically swept in steps of two degrees over a spherical section spanning elevation angles $\pm30^{\circ}$ and azimuth angles $\pm45^{\circ}$ to measure the omnidirectional phased array response. Fig. \ref{fig:bp_quantized} shows a heat map of the received power, where the color intensities of the hexagonal faces correspond to the power received at the beam pointing angles (the center of the faces). This indicates the measured signal strength vs. beam direction of all four phased arrays combined with their orientation on the receiver. Ideally, this heatmap should have a single hot spot in the middle pointing directly at the transmitter. However, due to the beam width and sidelobes, some energy is expected to be captured in other directions as well. Energy will also be captured by non-forward facing arrays due to leakage from the side and back walls of the array module. An example of such leakage can be seen at roughly $\pm90^{\circ}$ counterclockwise to the transmitter direction. This leakage is occurring due to reflections off the mounting platform used for chamber measurements. Additional work is needed to design RF absorbing covers for the mounting platform in order to suppress these unwanted reflections.


\section{Channel Measurements and Preliminary Results}\label{sec:measurements}
\subsection{Measurements}
The ROACH system was used to conduct both V2I and V2V channel measurements at the $28\,\rm{GHz}$ band. An experimental permit to temporarily transmit in this band was granted by the Federal Communications Commission (FCC)\cite{STA}.

\begin{figure}
 	\centering
 	\includegraphics[width=0.45\textwidth]{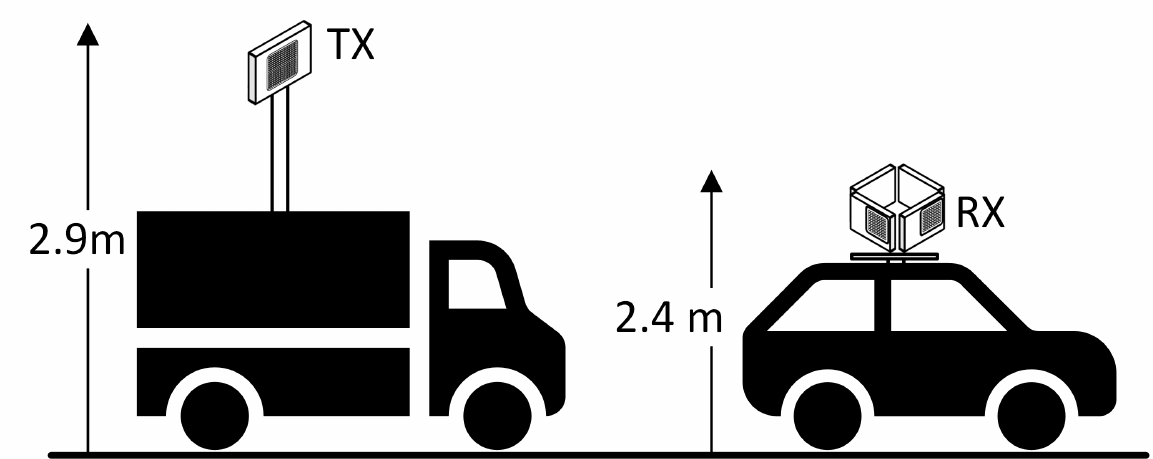}
 	\caption{The transmitter and receiver vehicles used for channel sounding campaigns.}
 	\label{fig:vehicles} 	
\end{figure}

\begin{figure}
	\centering
	\includegraphics[width=0.5\textwidth]{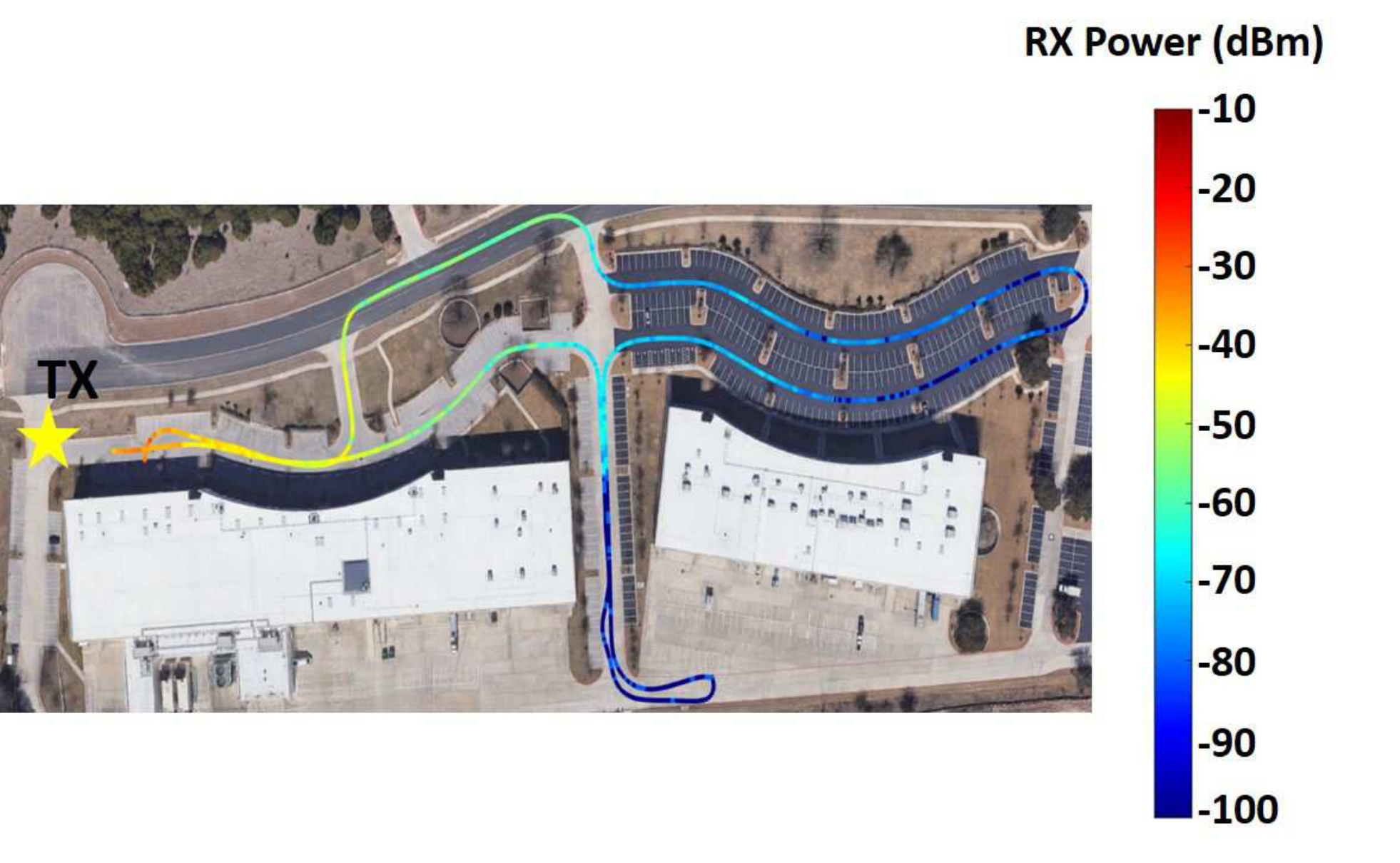}
	\caption{The received power by the best beam in Run 3, an exemplary Type II V2I run.}
	\label{fig:power_map} 	
\end{figure}

V2I channel measurements were conducted in and around two outdoor urban light commercial areas containing offices and retail, cars, and light pedestrian traffic. The first area was adjacent to multiple shopping complexes and office buildings, while the second area contained office complexes and residential apartment buidlings. In each of these areas the transmitter was either mounted on an accessible roof of a building at a height of $15\,\rm{m}$ (Type I measurement), or on an extendable mast on top of a stationary van at a height of $2.9\,\rm{m}$ as shown in Fig. \ref{fig:vehicles} (Type II measurement). In each case the height of the transmitter was recorded before collecting measurements. The receiver array system was mounted on the roof of a van, at a height of $2.4\,\rm{m}$. The receiver baseband system was placed in a rack mount system inside the van. The receiver van was driven around on surface roads and shopping and office parking lots at an average speed of $20\,\rm{mph}$ near the transmitter, ensuring that both line-of-sight and non line-of-sight locations were covered. The user interface would indicate when the received signal power was below measurement limits, and the driver would move the van towards a more favorable location. A map depicting the power received by the strongest beam in a sample Type II run is shown in Fig. \ref{fig:power_map}.

\begin{figure}
	\centering
	\includegraphics[width=0.5\textwidth]{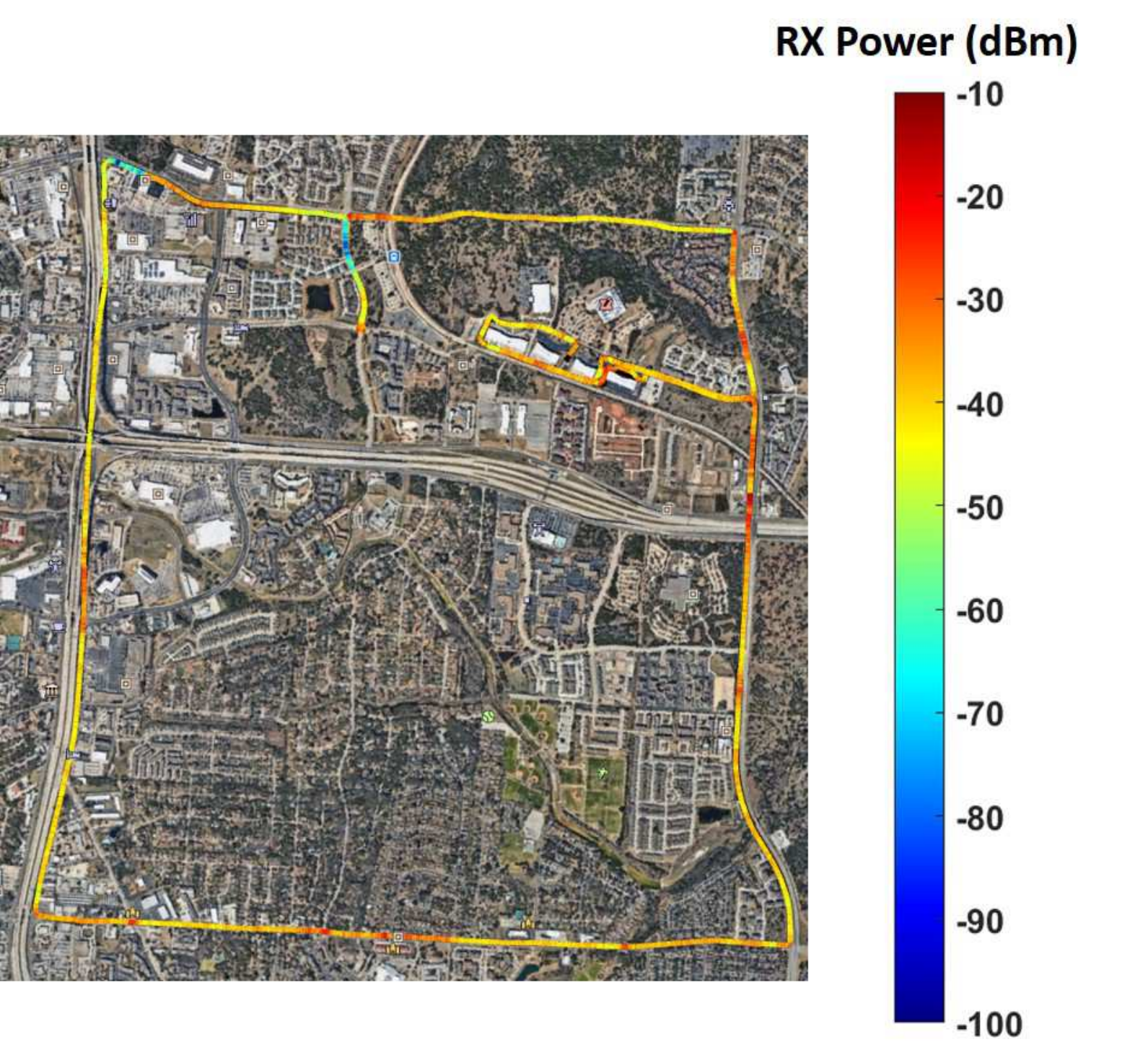}
	\caption{The received power by the best beam in an exemplary V2V run.}
	\label{fig:v2v_power_map} 	
\end{figure}

V2V measurements were also conducted in two outdoor areas covered by the experimental permit, where both the transmitter and receiver vans were in motion. For safety reasons, the transmitter mast was lowered to its base position while conducting mobile transmitter tests. Both vans were driven in light to moderate traffic and would allow other cars and trucks to freely move in or out between them. The trasnmitter and receiver separation would vary but typically maintained under 10 car lengths. The video captures that accompanied the channel measurements prove highly valuable as they allow correlation of non boresight signal reception with the location of nearby vehicles. A map depicting the power received by the strongest beam in a sample V2V run is shown in Fig. \ref{fig:v2v_power_map}.

\subsection{Preliminary Results} 

By analyzing the video recorded by the $360^{\circ}$ camera, line-of-sight (LOS) locations were identified during post-processing. The path loss experienced by a RX with ``perfect beam management" in LOS and NLOS scenarios was considered, i.e. the path loss experienced if beamforming at the RX is always done in the direction of maximum received power. Additionally, the ``boresight" LOS path loss was measured, equal to the power received by the beam pointing in the direction of the TX, with the RX heading calculated from GPS logs. During analysis, measurements were averaged over local squares with side lengths of $4\,\rm{m}$ to ameliorate the effects of small scale fading.  The measured pathloss was fit to the close-in (CI) pathloss model with $1\,\rm{m}$ reference distance\cite{Rappaport_2013b}. 
%
%
 \begin{figure}
	\centering
	\includegraphics[width=0.5\textwidth]{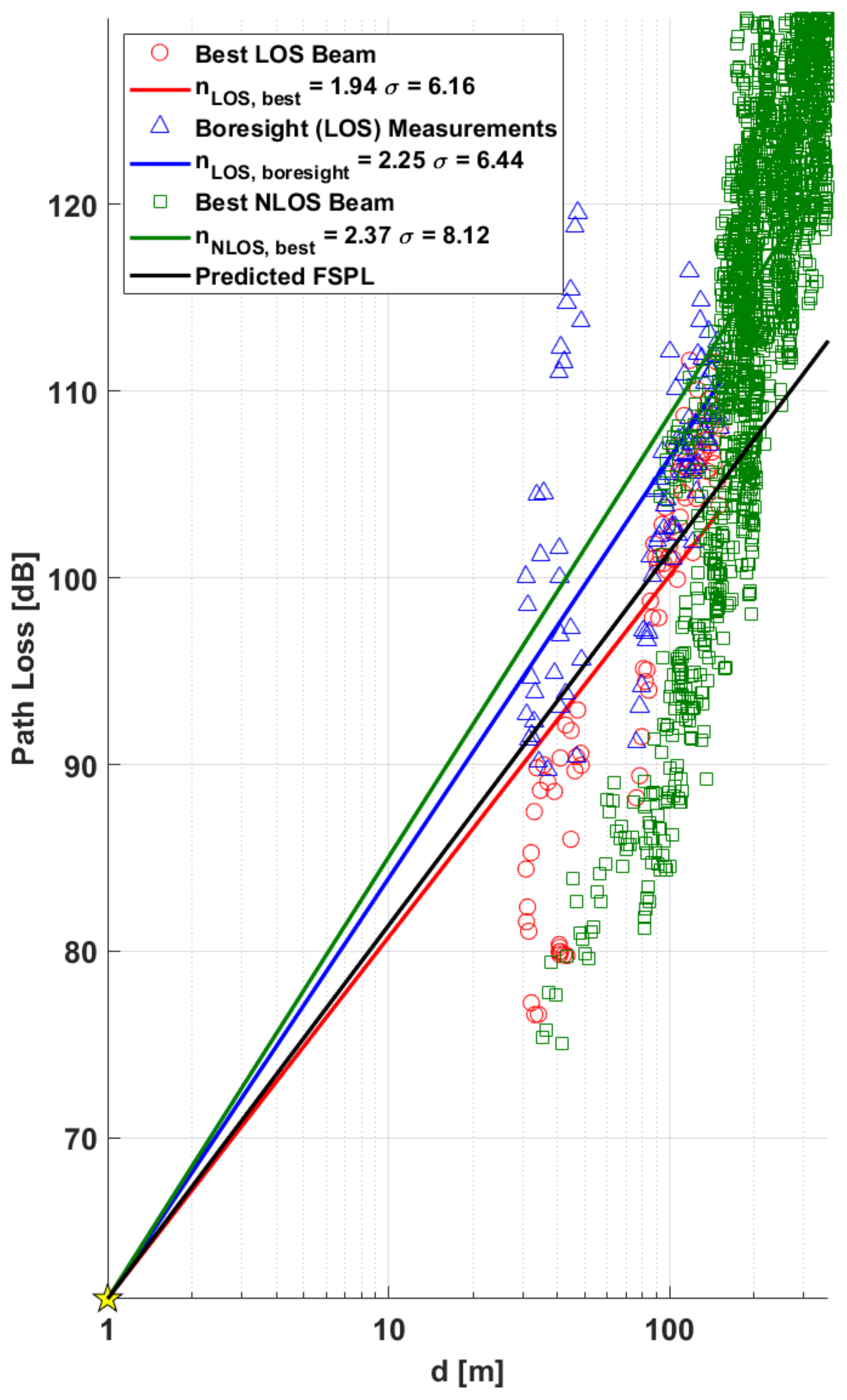}
	\caption{Scatter plots showing $28\,\rm{GHz}$ V2I pathloss vs. TX-RX separation for Type-I measurements. The lines indicate linear fit to the close-in pathloss model\cite{rural}.}
	\label{fig:path_loss_arb}	
\end{figure}
 \begin{figure}
	\centering
	\includegraphics[width=0.5\textwidth]{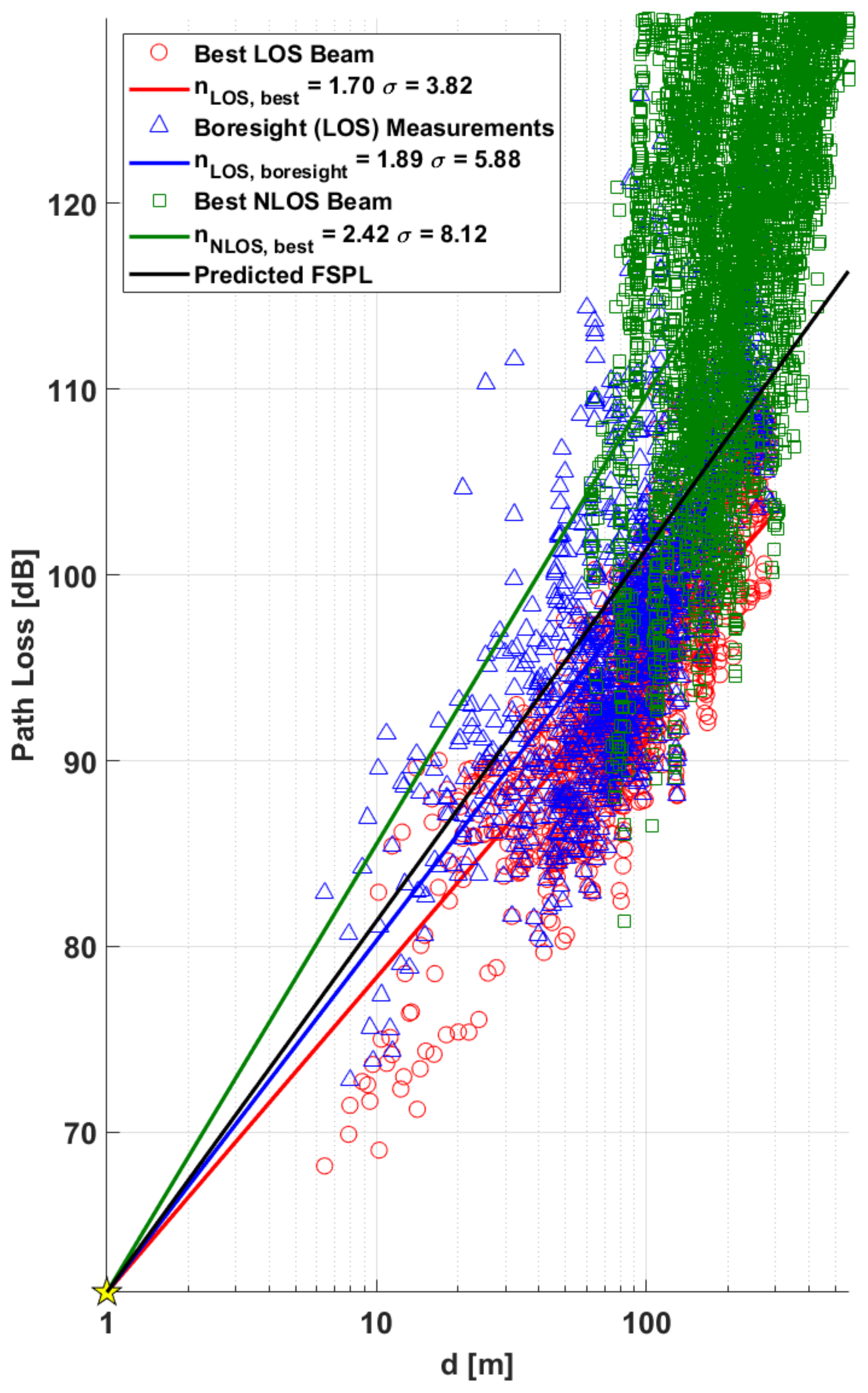}
	\caption{Scatter plots showing $28\,\rm{GHz}$ V2I pathloss vs. TX-RX separation for Type-II measurements. The lines indicate linear fit to the close-in pathloss model\cite{rural}.}
	\label{fig:path_loss_spect}	
\end{figure}
In Type I V2I measurements, where the TX was mounted on an accessible roof, a path loss exponent (PLE) of 1.94 was observed with the best LOS beam, as seen in Fig. \ref{fig:path_loss_arb} while a PLE of 1.70 over 5 test runs was observed for the best LOS beam in Type II V2I measurements, where the TX was mounted on a van at a height of 2.9 m, as seen in Fig. \ref{fig:path_loss_spect}. The PLE for each individual Type II run is listed in Table \ref{tbl:PLE_runs}. The PLE measured in Type II V2I measurements agrees with the PLE observed in static V2V LOS measurements conducted in \cite{Yamamoto_2008}, where a median PLE of 1.77 was observed. A PLE of 1.89 was observed over the Type II measurements when the RX pointing directly towards the TX (in boresight), which is close to the free space path loss (FSPL) exponent of 2.00 and is comparable to the PLE of 2.23 observed in \cite{Ben_Dor_2011}.


\begin{table}[h]
	\centering
	\caption{Pathloss Exponent Across All Type-II Runs  }
	\label{tbl:PLE_runs}
	\begin{tabular}{|l|l|l|l|l|l|l|}
		\hline
		& \multicolumn{2}{l|}{LOS, Best} & \multicolumn{2}{l|}{LOS, Boresight} & \multicolumn{2}{l|}{NLOS, Best} \\ \hline
		Run & n            & $\sigma (\rm{dB})$        & n              & $\sigma (\rm{dB})$           & n            & $\sigma (\rm{dB})$         \\ \hline
		1   & 1.79         & 3.21            & 1.90           & 4.07               & 2.25         & 8.92             \\ \hline
		2   & 1.66         & 3.67            & 1.88           & 6.14               & 2.37         & 6.00             \\ \hline
		3   & 1.57         & 2.16            & 1.79           & 4.92               & 2.53         & 3.42             \\ \hline
		4   & 1.59         & 3.84            & 1.92           & 7.16               & 2.67         & 9.15             \\ \hline
		5   & 1.67         & 3.79            & 1.95           & 7.46               & 2.40         & 9.12             \\ \hline
	\end{tabular}
\end{table}

\begin{figure}[h]
	\centering
	\includegraphics[width=0.5\textwidth]{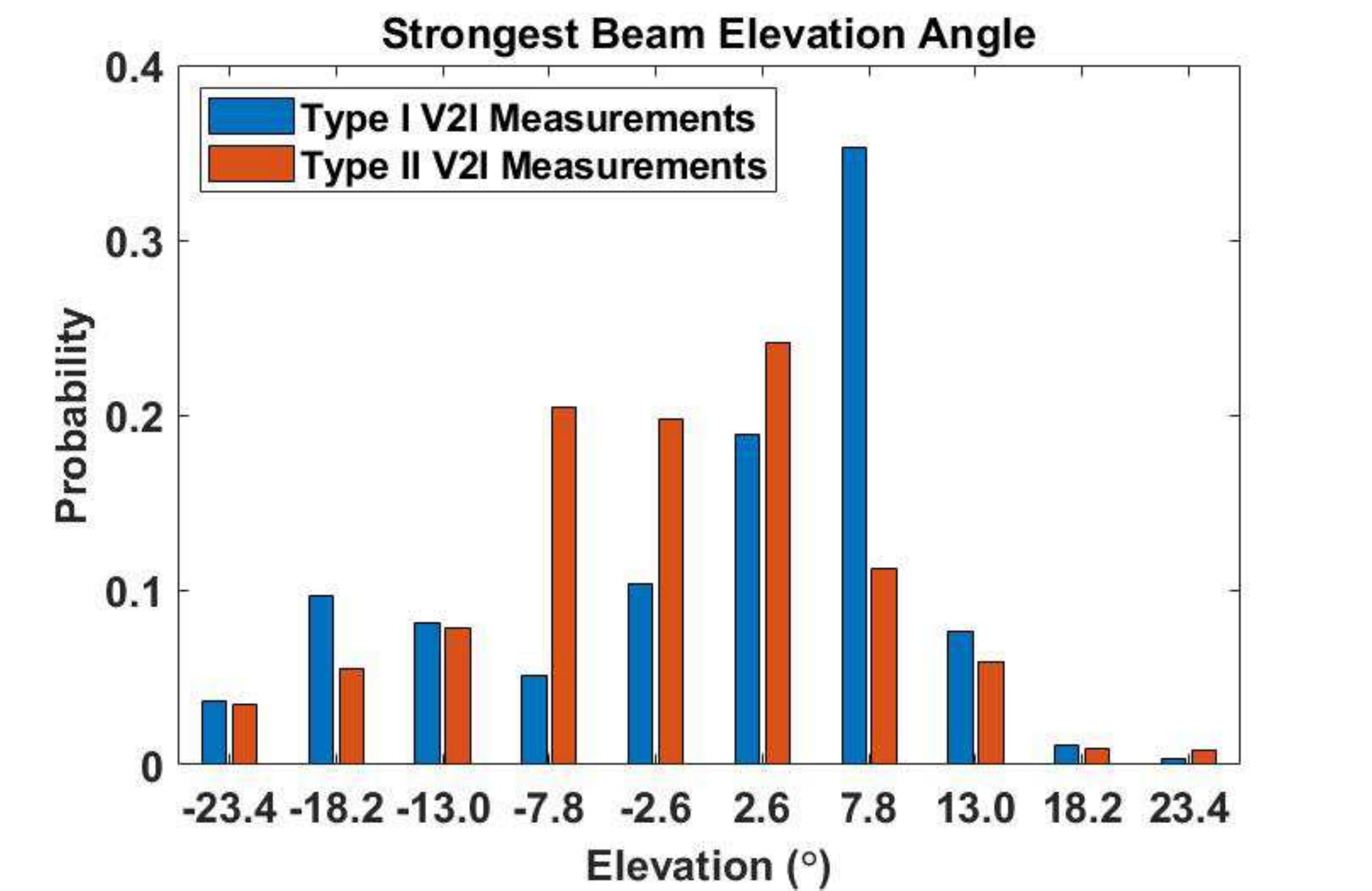}
	\caption{Histogram of the elevation angle of the strongest received beam measured in Type I and Type II V2I sounding campaigns.}
	\label{fig:elevation_angles_histogram}
\end{figure}

The low PLE observed in the direction of the strongest beam, particularly in Type II V2I measurements, suggest the presence of strong reflections. As seen in the histogram displayed in Fig. \ref{fig:elevation_angles_histogram}, for Type II measurements, the strongest beam elevation angle is the boresight angle ($ \pm 2.6^\circ $) only 47\% of the time, while the strongest signal is received at an angle of $-7.8^{\circ}$ 20\% of the time, which suggests reflections off metallic car roofs are an important propagation path.
 
\section{Conclusions and Future Work}\label{sec:conclusion}
In this paper we have introduced the design and architecture of ROACH, a novel real-time omnidirectional mmWave channel sounder capable of conducting real-time channel measurements at $28\,\rm{GHz}$ or $39\,\rm{GHz}$, with bandwidth up to $80\,\rm{MHz}$, a 200 beam omnidirectional scan in $6.25\,\rm{ms}$, with a mobile transmitter and mobile receiver, sub-1m accurate synchronized GPS position logging, and accompanied 8K resolution $360^{\circ}$ video. The system calibration procedure and the OTA verification procedure is described along with the measurement setup for V2I and V2V outdoor channel sounding campaigns. The preliminary path loss measurement and directionality analysis from multiple channel sounding campaigns is also provided. The analysis demonstrates the presence of multiple strong reflecting paths in a V2I campaign in and around an urban and suburban environment.

The next generation of ROACH currently under development has higher channel sounding bandwidth in order to achieve finer delay resolution of channel impulse response measurements. The phased array assembly is also being modified to fit on vehicle bumpers, which is a more likely location for vehicular antenna placement. Further analysis on the impact of beam management on channel statistics from the collected data is also left for future work. 

\bibliographystyle{IEEEtran}
\bibliography{references}{}

\begin{thebibliography}{10}
\providecommand{\url}[1]{#1}
\csname url@samestyle\endcsname
\providecommand{\newblock}{\relax}
\providecommand{\bibinfo}[2]{#2}
\providecommand{\BIBentrySTDinterwordspacing}{\spaceskip=0pt\relax}
\providecommand{\BIBentryALTinterwordstretchfactor}{4}
\providecommand{\BIBentryALTinterwordspacing}{\spaceskip=\fontdimen2\font plus
\BIBentryALTinterwordstretchfactor\fontdimen3\font minus
  \fontdimen4\font\relax}
\providecommand{\BIBforeignlanguage}[2]{{%
\expandafter\ifx\csname l@#1\endcsname\relax
\typeout{** WARNING: IEEEtran.bst: No hyphenation pattern has been}%
\typeout{** loaded for the language `#1'. Using the pattern for}%
\typeout{** the default language instead.}%
\else
\language=\csname l@#1\endcsname
\fi
#2}}
\providecommand{\BIBdecl}{\relax}
\BIBdecl

\bibitem{ATT_mmWave}
``{AT\&T Enhances Spectrum Position Following FCC Auction 102},''
  \url{https://about.att.com/story/2019/att_enhances_spectrum_position.html},
  accessed: 2020-05-23.

\bibitem{Rappaport_2013b}
T.~S. Rappaport \emph{et~al.}, ``{Millimeter Wave Mobile Communications for
  {5G} Cellular: It Will Work!}'' \emph{IEEE Access}, vol.~1, pp. 335--349, May
  2013.

\bibitem{Va_2016}
V.~Va, T.~Shimizu, G.~Bansal, and R.~W. {Heath Jr.}, ``{Millimeter Wave
  Vehicular Communications: A Survey},'' \emph{Found. Trends{\textregistered}
  Netw.}, vol.~10, no.~1, pp. 1--113, 2016.

\bibitem{Rappaport_2019}
T.~S. Rappaport \emph{et~al.}, ``{Wireless Communications and Applications
  Above 100 GHz: Opportunities and Challenges for 6G and Beyond},'' \emph{IEEE
  Access}, vol.~7, pp. 78\,729--78\,757, June 2019.

\bibitem{rappaport_2015}
T.~Rappaport, R.~Heath, R.~Daniels, and J.~Murdock,
  \emph{\BIBforeignlanguage{English (US)}{Millimeter wave wireless
  communications}}.\hskip 1em plus 0.5em minus 0.4em\relax Prentice Hall, 2015,
  includes bibliographical references (pages 585-651) and index.

\bibitem{MacCartney_2015}
G.~R. {MacCartney, Jr.} and T.~S. Rappaport, ``{A Flexible Millimeter-Wave
  Channel Sounder with Absolute Timing},'' \emph{IEEE Journal on Selected Areas
  in Communications}, vol.~35, no.~6, pp. 1402--1418, June 2017.

\bibitem{Park_2018}
J.~{Park}, J.~{Lee}, J.~{Liang}, K.~{Kim}, K.~{Lee}, and M.~{Kim},
  ``{Millimeter Wave Vehicular Blockage Characteristics Based on 28 GHz
  Measurements},'' in \emph{2017 IEEE 86th Vehicular Technology Conference
  (VTC-Fall)}, Feb. 2017, pp. 1--5.

\bibitem{Prokes_2018}
A.~{Prokes}, J.~{Vychodil}, T.~{Mikulasek}, J.~{Blumenstein}, E.~{Zöchmann},
  H.~{Groll}, C.~F. {Mecklenbräuker}, M.~{Hofer}, D.~{Löschenbrand},
  L.~{Bernadó}, T.~{Zemen}, S.~{Sangodoyin}, and A.~{Molisch}, ``{Time-Domain
  Broadband 60 GHz Channel Sounder for Vehicle-to-Vehicle Channel
  Measurement},'' in \emph{2018 IEEE Vehicular Networking Conference (VNC)},
  Dec. 2018, pp. 1--7.

\bibitem{Ben_Dor_2011}
E.~Ben-Dor, T.~S. Rappaport, Y.~Qiao, and S.~J. Lauffenburger,
  ``{Millimeter-wave 60 GHz outdoor and vehicle AOA propagation measurements
  using a broadband channel sounder},'' \emph{GLOBECOM - IEEE Glob. Telecommun.
  Conf.}, pp. 8--13, 2011.

\bibitem{Bas_2019}
C.~U. {Bas}, R.~{Wang}, S.~{Sangodoyin}, D.~{Psychoudakis}, T.~{Henige},
  R.~{Monroe}, J.~{Park}, C.~J. {Zhang}, and A.~F. {Molisch}, ``{Real-Time
  Millimeter-Wave MIMO Channel Sounder for Dynamic Directional Measurements},''
  \emph{IEEE Transactions on Vehicular Technology}, vol.~68, no.~9, pp.
  8775--8789, July 2019.

\bibitem{Caudill_2019}
D.~{Caudill}, P.~B. {Papazian}, C.~{Gentile}, J.~{Chuang}, and N.~{Golmie},
  ``Omnidirectional channel sounder with phased-arrayantennas for 5g mobile
  communications,'' \emph{IEEE Transactions on Microwave Theory and
  Techniques}, vol.~67, no.~7, pp. 2936--2945, 2019.

\bibitem{silva_2018}
M.~M. Silva, L.~da~Silva~Mello, P.~G. Castellanos \emph{et~al.}, ``Wideband
  channel sounding using modulated ofdm signals,'' in \emph{2018 IEEE-APS
  Topical Conference on Antennas and Propagation in Wireless Communications
  (APWC)}, 2018, pp. 1--4.

\bibitem{hua_2014}
M.~Hua, M.~Wang, K.~W. Yang, and K.~J. Zou, ``Analysis of the frequency offset
  effect on zadoff--chu sequence timing performance,'' \emph{IEEE Transactions
  on Communications}, vol.~62, no.~11, pp. 4024--4039, 2014.

\bibitem{emlid}
\BIBentryALTinterwordspacing
``Reach m2/m – rtk/ppk gnss modules for high precision mapping.'' [Online].
  Available: \url{https://emlid.com/reach/}
\BIBentrySTDinterwordspacing

\bibitem{smartnet}
\BIBentryALTinterwordspacing
``Smartnet north america.'' [Online]. Available:
  \url{https://www.smartnetna.com/resources_corrections.cfm}
\BIBentrySTDinterwordspacing

\bibitem{usrp}
\BIBentryALTinterwordspacing
``Getting started guide usrp-2950/2952/2953/2954/2955.'' [Online]. Available:
  \url{https://www.ni.com/pdf/manuals/376355c.pdf}
\BIBentrySTDinterwordspacing

\bibitem{3GPP.38.211}
3GPP, ``{5G; NR; Physical channels and modulation (Release 15)},'' TS 38.211
  V15.8.0, Jan. 2020.

\bibitem{STA}
``{Federal Communications Commission Experimental Radio Station Permit and
  License},'' \url{https://apps.fcc.gov/els/GetAtt.html?id=235534&x=.},
  accessed: 2020-05-23.

\bibitem{rural}
G.~R. {MacCartney} and T.~S. {Rappaport}, ``Rural macrocell path loss models
  for millimeter wave wireless communications,'' \emph{IEEE Journal on Selected
  Areas in Communications}, vol.~35, no.~7, pp. 1663--1677, 2017.

\bibitem{Yamamoto_2008}
A.~Yamamoto, K.~Ogawa, T.~Horimatsu, A.~Kato, and M.~Fujise, ``{Path-loss
  prediction models for intervehicle communication at 60 GHz},'' \emph{IEEE
  Trans. Veh. Technol.}, vol.~57, no.~1, pp. 65--78, Jan. 2008.

\end{thebibliography}

\end{document}